\documentclass[paper]{JHEP3}
\usepackage{epsfig}

\title{Precision tests of QED and non-standard models
by searching photon-photon scattering in vacuum with high power lasers}

\author{Daniele Tommasini\\ Departamento de F\'\i sica Aplicada, Universidade de Vigo.
As Lagoas E-32004 Ourense, Spain\\ E-mail: \email{daniele@uvigo.es}}

\author{Albert Ferrando\\ Interdisciplinary Modeling Group, InterTech. Departament d'\`{O}ptica,
Universitat de Val\`{e}ncia. Dr. Moliner, 50. E-46100 Burjassot (Val\`{e}ncia),
Spain.}

\author{Humberto Michinel\\ Departamento de F\'\i sica Aplicada, Universidade de Vigo.
As Lagoas E-32004 Ourense, Spain} 

\author{Marcos Seco\\ Departamento de F\'{\i}sica de Part\'{\i}culas, Universidade de Santiago de Compostela.
15706 Santiago de Compostela, Spain.}

\preprint{J. High Energy Phys. 11 (2009) 043}

\abstract{
We study how to search for photon-photon scattering in vacuum 
at present petawatt laser facilities such as HERCULES, 
and test Quantum Electrodynamics and non-standard models like Born-Infeld theory
or scenarios involving minicharged particles or axion-like 
bosons.
First, we compute the phase shift that is produced 
when an ultra-intense laser beam 
crosses a low power beam, in the case 
of arbitrary polarisations.
This result is then used in order to design a complete test 
of all the parameters appearing
in the low energy effective photonic Lagrangian.
In fact, we propose a set of experiments that can be performed at 
HERCULES, eventually allowing either to detect 
photon-photon scattering as due to new physics, 
or to set new limits on the relevant parameters, 
improving by several orders of magnitude the current constraints 
obtained recently by PVLAS collaboration. 
We also describe a multi-cross optical mechanism
that can further enhance the sensitivity, enabling HERCULES to detect
photon-photon scattering even at a rate as small as that predicted by QED.
Finally, we discuss how these results can be improved at future 
exawatt facilities such as ELI, thus providing a new class
of precision tests of the Standard Model and beyond.
}


\begin{document}

\section{Introduction}

Photon-Photon Scattering (PPS) in vacuum is a still unconfirmed 
prediction of both Quantum Electrodynamics (QED)~\cite{Costantini1971}
and non-standard models such as Born Infeld theory~\cite{BornInfeld,Denisov}.
Additional contributions to the process can also appear 
in new physics scenarios involving minicharged~\cite{MCP} 
or axion-like~\cite{ALP} particles.
However, all the experiments that have been performed by now 
could only be used to set upper limits on the photon-photon 
cross section $\sigma_{\gamma\gamma}$. 
The best constraints were obtained recently by PVLAS collaboration~\cite{PVLAS}, 
and are still seven orders of magnitude above the QED prediction for 
$\sigma_{\gamma\gamma}$.
In the last few years, there has been an increasing interest in 
studying the possibility to detect PPS at future facilities,
using two possible strategies.
On one hand, the cross section for the process will be maximum
at a possible future photon-photon
collider~\cite{phcollider}, based on a free electron laser
producing two beams of photons in the MeV range.
A second approach will be to perform experiments at optical 
wavelengths, and compensate the smaller cross sections with 
a very high density and/or a long path of interaction of the colliding 
photons~\cite{PPSsearch,ourfirsttwo,nonlinprop,Zavattini-Calloni}.
Most of these proposals require the construction of new facilities,
that will eventually be available in the future, such as a free electron
laser and/or an exawatt laser. Two exceptions are Refs.
~\cite{nonlinprop,Zavattini-Calloni}, that discussed the possibility
of performing experiments at present facilities to improve the PVLAS limit
on the photon-photon cross section. 
However, even in these cases the predicted sensitivity 
was found to be sufficient to detect PPS of QED origin only at future
facilities, such as ELI~\cite{ELI} or VIRGO+~\cite{VIRGOplus}, respectively.

Here, we will propose a set of experiments that can already be performed 
at present petawatt laser facilities, such as HERCULES~\cite{mourou06,HERCULES}.
First, we perform a new theoretical computation to get a quantitative expression for 
the phase shift that is produced by PPS when two 
orthogonally polarised beams cross each other.
This result turns out to provide a test for PPS of QED origin
which is significantly more sensitive than our previous 
proposal~\cite{ourfirsttwo,nonlinprop}, 
in which the two crossing beams had the same polarisation.
Moreover, it can be used in combination with our previous result
to provide a full test of all the parameters appearing
in the low energy effective Lagrangian describing the photons
in non-standard models, such as Born Infeld theory 
and scenarios involving
minicharged (MCP) or axion-like (ALP) particles.
In fact, taking into account the precision that can be achieved
in the measurement of optical phase shifts or ellipticities~\cite{kang97,wise,PVLAS},
we propose a set of experiments that will allow either 
to detect PPS at HERCULES, 
or to set new limits on the relevant parameters, 
improving by several orders of magnitude the current constraints 
obtained by the PVLAS collaboration. 
We then propose a multi-cross optical mechanism
that can further improve the sensitivity 
of this set of experiments, 
eventually enabling HERCULES to detect PPS as predicted by QED.
Finally, we discuss how these results can be improved at future 
exawatt facilities such as ELI, thus providing a new class
of precision tests of the Standard Model and beyond.

\section{The effective Lagrangian for the electromagnetic fields 
in QED and in non-standard models}

We will consider the case of photon energies well below the
threshold for the production of electron-positron pairs,
and assume an effective Lagrangian for the electromagnetic fields 
${\bf E }$ and ${\bf B}$ of the form
\begin{equation}
{\cal L}={\cal L}_0
+\xi_L {\cal L}_0^2+ \frac{7}{4}\xi_T {\cal G}^2,
\label{L}
\end{equation}
being ${\cal L}_0=\frac{\epsilon_0}{2}\left({\bf E}^2-{c^2\bf B}^2\right)$
the Lagrangian density of the linear theory, 
${\cal G}=\epsilon_0 c({\bf E}\cdot {\bf B})$ 
and $\epsilon_0$ and $c$ the dielectric
constant and the speed of light in vacuum, respectively. 
The additional, non-linear terms, that appear multiplying the parameters
$\xi_L$ and $\xi_T$ in equation~(\ref{L}), are the only two
Lorentz-covariant terms that can be formed with the electromagnetic fields
at the lowest order above ${\cal L}_0$. Therefore, they will 
appear as the first correction
to the linear evolution both in QED and in non-standard models.

In fact, in QED photons can interact with each other through the interchange
of virtual charged particles running in loop box diagrams~\cite{Costantini1971}.
Besides other interesting effects~\cite{tommasini0203}, such an interaction
leads to the Euler-Heisenberg effective Lagrangian density~\cite{Euler-Heisenberg}, 
that coincides with equation~(\ref{L}) with the identification
$\xi_L^{QED}=\xi_T^{QED}\equiv \xi$, being
\begin{equation}
\xi=\frac{8 \alpha^2 \hbar^3}{45 m_e^4 c^5}\simeq 6.7\times
10^{-30}\frac{m^3}{J}.
\label{constant_xi}
\end{equation}

On the other hand, in Born-Infeld theory~\cite{BornInfeld}, 
we would obtain the relation $\xi_T^{BI}=4\xi_L^{BI}/7$~\cite{Denisov}, 
in general without a definite prediction for the numerical value.

The presence of a minicharged (or milli-charged) particle (MCP)~\cite{MCP}
would provide an additional contribution analogous to that from the electron-positron
box diagram. If the new MCP are spin 1/2 fermions, and assuming that their mass
$m_\epsilon$ is still larger than the energy of the photons (the eV scale in optical
experiments), we would obtain
\begin{equation}
 \Delta\xi_L^{\rm MCP}=\Delta\xi_T^{\rm MCP}=
\left(\frac{\epsilon\, m_e}{m_\epsilon}\right)^ 4 \xi,
\end{equation}
where $\epsilon$ is the ratio of the charge of the particle with respect to
the electron charge. 
The existing laboratory bounds in this regime is 
$\epsilon\lesssim 8\times 10^ {-5}$~\cite{MCP}.
Taking masses above the eV scale, 
in order to apply the effective Lagrangian approach, this 
limits can be read as $\Delta\xi_L^{\rm MCP}\lesssim o(10^{6}\xi)$.
As we shall see in the next sections, this constraint has been improved by
PVLAS collaboration, and can be further strengthened by the experiments that
we propose in the present paper.
Of course, in this case there are already stronger
limits, $\epsilon\lesssim 10^ {-15}$, from astrophysical and 
cosmological observations~\cite{MCP}.
A larger contribution might be obtained if the MCP
are lighter than the energy scale of the photons (the eV scale in the present paper). 
However, this case would deserve
a different treatment which goes beyond the purposes of the present
work, since it cannot be described simply by an effective Lagrangian
of the form of equation (\ref{L}). 

Similar considerations apply if the new MCP is a spinless boson. 
Assuming again that its mass
$m_\epsilon$ is still larger than the energy of the photons (the eV scale in optical
experiments), and using the results of Ref.~\cite{Kruglov}, we would obtain
\begin{equation}
 \Delta\xi_L^{\rm MCP0}=\frac{7}{16}\left(\frac{\epsilon\, m_e}{m_\epsilon}\right)^ 4 \xi
\end{equation}
and 
\begin{equation}
\Delta\xi_T^{\rm MCP0}=\frac{1}{28}\left(\frac{\epsilon\, m_e}{m_\epsilon}\right)^ 4 \xi. 
\end{equation}
On the other hand, if the MCP is a spin 1 boson, the contribution to the 
effective Lagrangian would be larger, as computed using 
the result of Refs.~\cite{Kruglov}. We obtain
\begin{equation}
 \Delta\xi_L^{\rm MCP1}=\frac{261}{16}\left(\frac{\epsilon\, m_e}{m_\epsilon}\right)^ 4 \xi
\end{equation}
and
\begin{equation}
 \Delta\xi_T^{\rm MCP1}=\frac{243}{28}\left(\frac{\epsilon\, m_e}{m_\epsilon}\right)^ 4 \xi
\end{equation}

Let us now discuss the case of an axion-like particle~\cite{ALP}. This can be 
a Light Pseudoscalar Boson or a Light Scalar Boson, 
depending on the coupling with the photons,
that is described in the Lagrangian density by the terms 
${\cal L}_P= -\sqrt{\hbar\, c}\,g_P \Phi_P {\cal G}$ 
and ${\cal L}_S= -\sqrt{\hbar\, c}\,g_S \Phi_S {\cal L}_0$,
respectively.
We can find the leading contribution to the effective Lagrangian when
the photon energy is much smaller than the $m_\Phi$ scale, that can be cast 
in the form of equation~(\ref{L}) with an additional contribution given by
\begin{equation}
 \Delta\xi_T=\frac{2\hbar^3g_P^2}{7 c\, m_\Phi^2} 
\label{xi_Pseudoscalar}
\end{equation}
and $\Delta \xi_L=0$, in the case of pseudoscalars, or 
\begin{equation}
 \Delta\xi_L=\frac{\hbar^3g_S^2}{2 c\, m_\Phi^2}
\label{xi_Scalar}
\end{equation}
and $\Delta \xi_T=0$, in the case of scalars.
On the other hand, for $m_\Phi\gtrsim 1{\rm eV}$, the Cristal Ball \cite{ALP}
laboratory limit
$g_P\le 4.2\times 10^{-3}{\rm GeV}^{-1}$ gives the contraint
$g_P/(m_\Phi c^2)\lesssim 4\times 10^{6}{\rm GeV}^{-2}$, which can be converted in 
the limit $\Delta \xi_T\lesssim2.2\times10^{-25}m^3/J$.
This constraint has been improved recently 
by the PVLAS consideration~\cite{PVLAS}, as we shall see 
in the next section. Again, the astrophysical limits $g_P\lesssim2.7\times10^{-9}{\rm GeV}^{-1}$,
valid for $m_\Phi\lesssim1 {\rm KeV}$ ~\cite{ALP}, 
is still much more stringent than any laboratory bound.

Similar considerations apply for scalar boson, for which 
the best laboratory constraints are also those that were recently
set by PVLAS, that we will review in the next section.
We also recall that our approximations do not apply 
for masses smaller than the order of the energy of the 
colliding photons. In this case, the computation of $\Delta\xi_L$
is more complicated, and the production of real axions has also to be taken into 
account. The latter can be expected to produce 
dichroism, just as in the presence of a constant external magnetic 
field~\cite{magnetic_bire,PVLAS},
and combined with the measurement of ellipticity may allow for a determination
of both $m_\Phi$ and $g_{S,P}$. However, this case lies beyond the scope
of the present paper, that uses a phenomenological approach
that can be applied to any theory that goes beyond the 
Standard Model, in the energy regime in which it only implies 
a different contribution to equation (\ref{L}), as parametized by artitrary 
$\xi_L$ and $\xi_T$.
Expressing the electromagnetic fields in terms of the four-component gauge
field $A^\mu=(A^0,{\bf A})$ as ${\bf B}=\nabla \wedge {\bf A}$ and
${\bf E}=-c \nabla A^0-\frac{\partial {\bf A}}{\partial t}$, 
this gives the equations of motion as the variational derivatives
${\delta \Gamma}/{\delta A^\mu}=0$,
where $\Gamma\equiv \int {\cal L} d^4x$ is the effective
action. 
Such equations are similar to the modified nonlinear
Maxwell's equations that have been obtained in Ref.~\cite{mckenna},
the only difference being the distinction between $\xi_L$ and $\xi_T$.

\section{Present constraints}

The current limits on PPS in vacuum have been obtained recently by the PVLAS
collaboration~\cite{PVLAS} by searching evidence of 
birefringence of the vacuum in a uniform magnetic field background~\cite{magnetic_bire}.
Their negative result was used to set the current constraints on the parameters appearing in 
equation~(\ref{L}). 
With our notation, their 95 \% C.L. limit reads
\begin{equation}
 \frac{\vert 7\xi_T-4\xi_L\vert}{3}<3.2\times10^{-26}\frac{m^3}{ J}.
\end{equation}
Assuming $\xi_L=\xi_T\equiv\xi^{exp}$ as in QED, 
their results can be translated in the limit
$\xi^{exp}<3.2 \times 10^{-26}{m^3}/{J}$, which is
$4.6\times 10^{3}$ times higher than the QED value of equation~(\ref{constant_xi})
(7 orders of magnitude for the cross section $\sigma_{\gamma\gamma}$).

Note however that PVLAS experiment was only sensitive to the combination 
$\vert 7\xi_T-4\xi_L\vert$ of the parameters. In particular, 
this quantity vanishes when
$\xi_T=\frac{4}{7}\xi_L$, therefore PVLAS experiment is unable to set any constraint
on a pure Born Infeld theory.

\section{Approximate solution for the scattering of orthogonally polarised beams}

In Ref.~\cite{ourfirsttwo}, we have studied the 
scattering of two counter-propagating waves that are polarised 
in the same direction, and we have found that the effect of
PPS was to produce 
a phase shift in each wave, which was proportional to $\xi_L$ multiplied by the
intensity of the other wave. That result was obtained by an
analytical, variational approximation, and was shown to agree 
with a numerical solution of the full non-linear equations,
that was also obtained in the second of Refs.~\cite{ourfirsttwo}. 

Here, we will apply a similar variational method to find a
solution for the problem of the scattering of two orthogonally polarised 
counter-propagating waves, one of which represents an ultra-high power beam. 
Let the low power and the high power waves be polarised in the $x$ and $y$ directions
respectively, so that their linear (free) evolution (neglecting photon-photon scattering) 
would be $A_x(t,z)^{\rm lin}=\alpha_0\cos(k z - \omega t)$ 
and $A_y(t,z)^{\rm lin}={\cal A}\cos(k z + \omega t+\varphi)$,
where $\omega=k c$ and we allow for an initial phase difference $\varphi$. 
Their energy density, when each of the two waves 
is taken alone, would be 
$\rho_x=\epsilon_0 \omega^2{\alpha_0}^2/2$ and 
$\rho_y\equiv \rho=\epsilon_0 \omega^2{\cal A}^2/2$, respectively. 
Hereafter, we will assume that $\rho_y\gg\rho_x$.
When these two waves are made to scatter, they will affect each other due to 
PPS, as described by the non-linear terms in equation~(\ref{L}).
First, we note that the assumption of 
no dependence on $x$ and $y$ of the fields guarantees that the condition $A_t=A_z=0$ 
is maintained by the non-linear evolution, 
since we have checked that in this case the equations 
${\delta \Gamma}/{\delta A_t}=0$ and ${\delta \Gamma}/{\delta A_z}=0$ 
are automatically satisfied, independently on the values of $A_x(t,z)$ and $A_y(t,z)$. 
Therefore, in the absence of $x$ and $y$ dependence, 
the components $A_t$ and $A_z$ with will 
not be generated if they are not present from the beginning.
Second, we note that the non-linear effect is driven by the very small 
parameters $\xi_L$ and $\xi_T$. This justifies a perturbatively-motivated 
variational approach, similar to that introduced in Refs.~\cite{ourfirsttwo}. 
We then need to chose a good ansatz for
the fields $A_x(t,z)$ and $A_y(t,z)$. 
In principle, each of the two components can get a transmitted wave contribution, 
propagating along the same direction as the original wave,
and a reflected wave propagating in the opposite direction. In a perturbative approach,
we can compute these different effects separately and then sum them up.
Therefore, we will first neglect the reflected waves,
and use the following ansatz:
\begin{eqnarray}
\label{var_orthogonalpol}
&A_x=\alpha(z)\cos(k z - \omega t)+\beta(z)\sin(k z - \omega t),&\\ \nonumber
&A_y={\cal A}\cos(k z + \omega t+\varphi).
\end{eqnarray}
Here, we have neglected the effect of the low power wave on the high power wave,
taking into account that such an effect is expected to be proportional to the
energy density of the low power beam. This expectation, inspired by our previous work 
~\cite{ourfirsttwo}, will be confirmed by the result that we will obtain below.

We now substitute the ansatz (\ref{var_orthogonalpol})
in the Lagrangian (\ref{L}), and average out the fast variation
in $z$ over distances of the order $2\pi /k$, assuming that the envelop functions 
$\alpha(z)$ and $\beta(z)$ will show a much slower variation, as we will verify a posteriori.
We then compute the variational equations 
$\delta\Gamma/\delta\alpha=0$ and $\delta\Gamma/\delta\beta=0$, 
keeping the lowest order terms in
the expansion parameter $\xi_T$ and neglecting the 
higher order space derivatives $(d/dz)^n$ of 
$\alpha(z)$ and $\beta(z)$, as compared with $k^n$. 
After a long but straightforward algebra, we find the following equations: 
\begin{eqnarray}
\label{var_equations}
&\beta'(z)+\chi_T \alpha(z)=0,&\\ \nonumber
&\alpha'(z)-\chi_T \beta(z)=0,
\end{eqnarray}
where $\chi_T\equiv 7 \epsilon_0 c^2 {\cal A}^2 k^3\xi_T/2$. 
Assuming the initial condition $\alpha(0)=\alpha_0$, $\beta(0)=0$,
in such a way that the corrected solution coincides initially with that
of the linear problem, we find then $\alpha(z)=\alpha_0 \cos(\chi_T z)$, 
$\beta(z)=-\alpha_0 \sin(\chi_T z)$.
After substituting in equation~(\ref{var_orthogonalpol}), 
we finally get the following variational solution:
\begin{eqnarray}
\label{varsol_orthogonalpol}
&A_x=\alpha_0\cos(k z + \chi_T z - \omega t),&\\ \nonumber
&A_y={\cal A}\cos(k z + \omega t+\varphi).
\end{eqnarray}

In other words, taking into account that $\epsilon_0{\cal A}^2 \omega^2/2\simeq\rho$ 
is the energy density of the high power wave,
we find that after a crossing distance $\Delta z$, 
the phase of the orthogonal, low power wave is shifted by an amount
$\Delta\phi_T=\chi_T \Delta z=7 \xi_T \rho k \Delta z$.
Note that this result is independent of the initial phase difference $\varphi$ 
between the two crossing waves.

Let us now introduce the possibility that a reflected wave is generated 
in the component $A_x$, as described by the ansatz
\begin{eqnarray}
 &A_x=\alpha_0\cos(k z +\chi_T z- \omega t)+\gamma(z)\cos(k z + \omega t)
 +\delta(z)\sin(k z + \omega t),&\\ \nonumber
 &A_y={\cal A}\cos(k z + \omega t+\varphi)
\end{eqnarray}
After repeating the same procedure as above, we get 
the following variational solution
\begin{eqnarray}
 &\gamma(z)=\frac{\alpha_0 k z}{4\pi} \left[\cos(14\pi\rho\xi_T)-1\right],&\\ \nonumber
 &\delta(z)=\frac{\alpha_0 k z}{4\pi} \sin(14\pi\rho\xi_T).
\end{eqnarray}
Now, the quantity $14\pi\rho\xi_T$ can be estimated 
for the petawatt laser HERCULES~\cite{HERCULES} that we will consider below
for our proposals of experiments. In this case, the peak intensity is 
$I\sim 2\times 10^{22}{ W}{ cm}^{-2}$~\cite{HERCULES},
corresponding to an energy density $\rho\sim 6.7\times 10^{17}{J}{m}^{-2}$.
Taking into account the PVLAS limit $\xi^{exp}<3 \times 10^{-26}{m^3}/{J}$,
and assuming that it can be applied to $\xi_T$ at least roughly (see also figure 2
in the last section), we find that $\xi_T\rho\lesssim 2 \times 10^{-8}$ for the
product giving the importance of the nonlinear QED effects.
As a result, the variational solution for $\gamma$ and $\delta$
implies that $\gamma(z)\simeq 0$ for all the practical purposes, and 
$\delta(z)\simeq 7 \alpha_0\rho \xi_T k z/2$. 
Taking $k\sim7.8\times10^{6}m^{-1}$ as for the wavelengths of e.g. the HERCULES 
laser ($\lambda=800nm$), we obtain that 
the $\vert\delta(z)/\alpha_0\vert\lesssim 0.5 z/1m$. 
Now even in the multi-cross configuration that will be discussed below
the crossing length will be smaller than the centimetre scale, so that
 $\delta(z)$ will be smaller than
the low power amplitude $\alpha_0$ at least 
by two orders of magnitude.  For this reason, it will be neglected.

Finally, let us introduce the possible reflected wave 
in the component $A_y$, as described by the ansatz
\begin{eqnarray}
 &A_x=\alpha_0\cos(k z +\chi_T z- \omega t),&\\ \nonumber
 &A_y={\cal A}\cos(k z + \omega t+\varphi)+\eta(z)\cos(k z - \omega t)+\sigma(z)\sin(k z - \omega t).
\end{eqnarray}
By repeating the same kind of computations and arguments as above, we find 
the following solution
\begin{eqnarray}
\label{varsol_total}
&A_x=\alpha_0\cos(k z +\chi_T z- \omega t),&\\ \nonumber
&A_y={\cal A}\cos(k z + \omega t+\varphi)+\eta_0\cos(k z + \chi_L z-\omega t),
\end{eqnarray}
where $\chi_L= 2 \epsilon_0 c^2 {\cal A}^2 k^3\xi_L=4\xi_L\rho k$,
$\eta(0)=\eta_0$ and we assume that $\sigma(0)=0$. 
We then see that the counter-propagating
wave in the $y$ polarisation only exists if it is present from the beginning,
and that it gets a phase shift $\Delta\phi_L=\chi_L\Delta z$ 
which is equal to that obtained in Ref.~\cite{ourfirsttwo},
as could be expected.

equation~(\ref{varsol_total}) implies that, 
after crossing a counter-propagating, linearly polarised ultra-intense laser pulse, 
an ordinary laser pulse is phase shifted both in the polarisations parallel and orthogonal
to that of the high power beam. The corresponding phase shifts are
\begin{eqnarray}
\label{phase_shift_xy}
&\Delta \phi_L=4 \xi_L\rho k \Delta z=4 \xi_L I k \tau,&\\ \nonumber
&\Delta \phi_T=7 \xi_T\rho k \Delta z=7 \xi_T I k \tau,
\end{eqnarray}
where $I=\rho c$ is the intensity of the high power beam and $\tau=\Delta z/c$ is its 
time length. If we assume $\xi_L=\xi_T$ as in QED, we see that $\Delta\phi_T$ 
is more sensitive by a factor $7/4$ than $\Delta\phi_L$ to the effect of PPS. 
This is already an improvement with respect to Ref.~\cite{ourfirsttwo}.
Moreover, the dependence of equations~(\ref{phase_shift_xy}) 
on both parameters $\xi_L$ and $\xi_T$
will permit a full analysis of the effective Lagrangian (\ref{L}),
distinguishing between QED and other models such as Born Infeld theories.
Finally, we note that (\ref{phase_shift_xy}) also implies that the high power pulse
behaves like a birefringent medium, producing a relative phase shift
$ \Delta\phi_b=\Delta\phi_T-\Delta\phi_L=(7\xi_T-4\xi_L) I k\tau$
between the transverse an parallel polarisations of the low power beam.

\section{Proposal of experiments}

We will now discuss how the result of equations~(\ref{phase_shift_xy})
can be used to search PPS in vacuum 
by measuring phase shifts and ellipticities.
In fact, Ref.~\cite{kang97,wise} provides a technique that allows for 
the measurement of phase shifts 
as small as $10^{-8}rad$, which is the noise limit~\cite{wise}.
This precision, that holds for ultra-short laser pulses~\cite{kang97,wise}, applies 
then to our $\Delta\phi_L$ and $\Delta\phi_T$.
A similar sensitivity can be obtained for the measurement of the ellipticity
induced by $\Delta\phi_b$ corresponding to birefringence. In particular, 
in the same experiment that we have cited above~\cite{PVLAS}, 
the PVLAS collaboration was able to resolve the corresponding $\Delta\phi_b$ 
with a statistical error $\sigma_b=1.1\times 10^{-8}rad$,
thus allowing them to set a 95\% C.L. experimental limit $\Delta\phi_b<2.8\times 10^{-8}rad$.
Hereafter, to be definite and for simplicity, we will use this same 
numerical value,
$2.8\times 10^{-8}rad$, for the sensitivity in 
the measurement of $\Delta\phi_L$ and $\Delta\phi_T$, taking into account
that the actual experimental precision will be close to this choice~\cite{kang97,wise}.

We can now propose a set of three experiments: 

1) A linearly polarised low power laser pulse
is divided by a beam splitter in two branches. One of them 
propagates freely in vacuum, while the other
crosses a contra-propagating ultra-high power laser pulse
polarised in the same direction.
The phase shift suffered by the low power pulse 
as a consequence of PPS
is then measured by comparing with the
pulse that has propagated freely, using the technique described in Ref.
~\cite{kang97,wise}. 
Due to equation~(\ref{phase_shift_xy}), this configuration 
can be used to measure the parameter 
$ \xi_L=\frac{\Delta\phi_L}{4 F I k \tau}$,
where we have introduced a gain factor $F$ that corresponds to a multi-cross 
configuration as discussed below. 
This experiment will then allow either to detect
PPS by measuring a non-vanishing $\xi_L$, or to 
set an upper limit on this parameter as
\begin{equation}
\label{limit_xi_L}
 \xi_L<\frac{2.8\times 10^{-8}}{4 F I k \tau}.
\end{equation}

2) The configuration is the same as in case 1), 
except that now the high power beam is polarised in a direction orthogonal to that
of the low power pulse. 
Due to equation~(\ref{phase_shift_xy}), this setup can be used to obtain the parameter 
$\xi_T=\frac{\Delta\phi_T}{7 F I k \tau}$ by measuring the phase shift $\Delta\phi_T$.
This will allow either to detect PPS or to set the upper limit 
\begin{equation}
\label{limit_xi_T}
 \xi_T<\frac{2.8\times 10^{-8}}{7 F I k \tau}.
\end{equation}

3) The low power beam polarisation has two components, one parallel 
and another orthogonal to that of the contra-propagating high power pulse. 
Ellipticity measurements can then be used to deduce the difference of the phase shifts
$\Delta\phi_T-\Delta\phi_L=\Delta\phi_b$, allowing to determine the
combination $\frac{7\xi_T-4\xi_L}{3}=\frac{\Delta\phi_b}{3 F I k \tau}$,
thus allowing either to detect PPS, or to 
set the upper limit 
\begin{equation}
\label{limit_xi_b}
 \frac{\vert 7\xi_T-4\xi_L\vert}{3}<\frac{2.8\times 10^{-8}}{3 F I k \tau}.
\end{equation}

The combination of these three experiments will permit a complete exploration
of the parameter space. Actually, it is easy to see that if $\xi_L$ and $\xi_T$
have the same sign, as in QED and Born Infeld theories, 
experiment 3) is less sensitive than the combination of the others and can be 
discarded without losing significant information. 
On the other hand, if $\xi_L$ and $\xi_T$
have opposite sign, experiment 3) is the most sensitive one, although even in this case the
other two measurements would be useful for a full determination of both
$\xi_L$ and $\xi_T$. 

From equations~(\ref{limit_xi_L}), (\ref{limit_xi_T}) and (\ref{limit_xi_b}), we see that 
the sensitivity depends on the combination $I k \tau$ of the
experimental parameters of the ultra-high power laser beam, and on a
gain factor $F$ that will be discussed later.
Therefore, the most favourable experimental configuration will be that
allowing for the maximum value of the product $I k \tau$.
As far as we know, the highest value achieved at present facilities is
that of HERCULES laser~\cite{HERCULES}, 
that reaches peak intensities $I=2\times 10^{22}W/cm^2$,
for a time length $\tau=3\times 10^{-14}s$ and wavelength 
$\lambda=2\pi/k=8.1\times 10^{-7}m$.
This gives $I k \tau=4.7\times10^{19}J/m^3$. 
Even in the absence of any gain factor ($F=1$),
such a facility will be able to resolve $\xi_L$ and $\xi_T$ as small as 
$1.5\times 10^{-28}m^3/J$ and $8.6\times 10^{-29}m^3/J$, respectively, 
thus allowing either to
detect PPS of non QED origin, or set limits on the parameters that are 
more than two orders of magnitude (5 order of magnitude in the cross section) 
more stringent than the current PVLAS limits,
in addition to the fact that they constrain the full parameter space, including
the case of Born-Infeld theories, that were unconstrained by PVLAS.

A significant improvement will be obtained in the near future at 
ELI~\cite{ELI}, that in its first stage 
will achieve peak intensities $I\simeq 10^{25}W/cm^2$,
for a time length $\tau\simeq 10^{-14}s$ and wavelength 
$\lambda=2\pi/k\simeq8\times 10^{-7}m$.
This gives $I k \tau\simeq 8\times10^{21}J/m^3$. Even in the absence of any gain factor ($F=1$),
such a facility will be able to resolve $\xi_L$ and $\xi_T$ as small as 
$9\times 10^{-31}m^3/J$ and $5\times 10^{-31}m^3/J$, respectively. 
In particular, ELI would allow for the 
detection of PPS of QED origin and for measuring the 
parameter $\xi$ with two figures.

\section{Improving the sensitivity with multiple crossing}

The sensitivity of our proposed
experiments can be enhanced by making the two beams cross each other several times,
using a kind of wave guide consisting of two parallel series of parabolic
mirrors as shown in figure 1. An advantage of using parabolic mirrors is that,
in the paraxial approximation, they do not generate aberrations in the beam.
We assume that the laser pulses are localised at a distance $R$ 
half a way to the path leading to the next mirror.
To be concrete, we will also assume that at the crossing points 
the high power laser is focused to the diameter $d\simeq 0.8\mu m$ 
and intensity $I\simeq 2\times10^{22}Wcm^{-2}$ of the HERCULES beam.
The time duration  $\tau=30fs$ implies that the pulse length $c\tau\simeq9\mu m$
along the direction of propagation
is approximately an order of magnitude greater than its transversal width, 
therefore the two beams must cross forming an angle 
$\theta$ close to $\pi$ in order to maximize their superposiposition.
These requirements may be achieved using plasma mirrors~\cite{Hora}, 
that can work e.g. at the intensity
$I_{\rm mirror}\simeq2\times10^{19}W/cm^2$ with a 
reflection coefficient $r\simeq0.98$~\cite{Hora}.
In this case, the high power beam at the mirror should have a diameter 
equal to $d\sqrt{I/I_{\rm mirror}}\simeq 25\mu m$. 
In order to avoid diffractive distortions, 
we will use parabolic mirrors of, say, a double diameter, 
$d_{\rm mirror}\simeq50 \mu m$.
On the other hand, the two planes are assumed to be at a distance $\simeq 2R$, with
$R$ much larger than $d_{\rm mirror}$, say $R=5cm$, in such a way that 
$\theta-\pi=2 \arccos(d_{\rm mirror}/R)-\pi\simeq-2\times10^{-3}$, so that $\theta$ 
is very close to $\pi$. A serious technological challenge to be faced 
will be the very precise alignment of the mirrors, since any uncertainty in the
direction will be multiplied by the number of times
the beams are reflected. In principle, the orientation of the mirrors can currently
be fixed with a precision as small as $\Delta\theta\sim10^{-8}rad$~\cite{angularprecision}. 
After $N$ reflections, this will produce an uncertainty $\sim N R\Delta\theta$ on the position 
of the spot at the focus. This uncertainty must be smaller than the diameter 
of the beams at the focus, so that $N\lesssim d/(R\Delta\theta)\sim10^3$.

%%%%%%%%%%%%%%%%%%%%%% FIGURE %%%%%%%%%%%%%%%%%%%
%\begin{figure}[htb]
%{\centering \resizebox{8.5cm}{!}{\includegraphics{multipass2.eps}} \par}
%\caption{Proposed setup for multiple crossing of the two scattering laser pulses.}
%\label{fig1}
%\end{figure}
%%%%%%%%%%%%%%%%%%%%%%%%%%%%%%%%%%%%%%%%%%%%%

\EPSFIGURE[r]{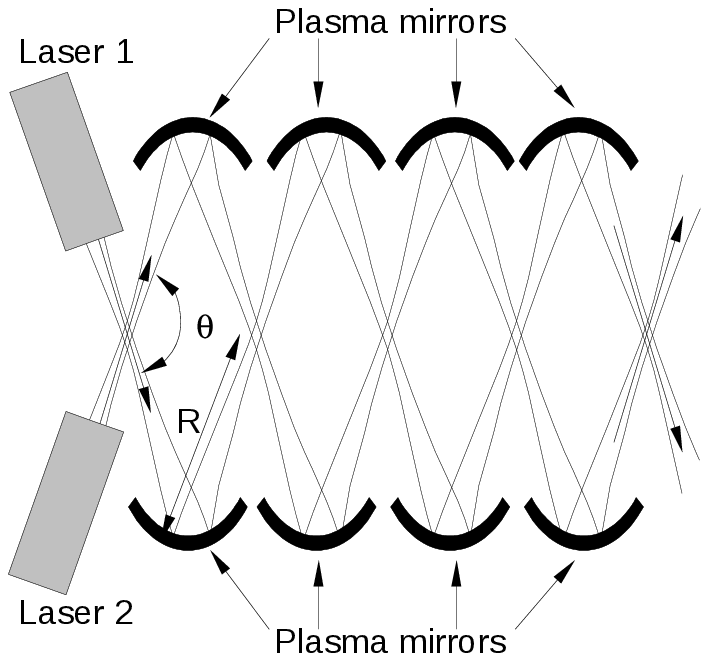}{Proposed setup for multiple crossing of the two scattering laser pulses.}

In order to compute the gain factor $F$ of this configuration, we note that 
after each reflection the intensity is reduced by a factor $r\simeq0.98$. Moreover, 
the phase shift is due to the counter-propagating components of the photon momenta,
$p_z=\hbar k\sin(\theta/2)$. Taking into account that $k$ appears to the third 
power in the expression of the phase shift (or equivalently that it appears
to the sixth power in the cross section~\cite{Costantini1971}), the gain factor 
is then $F=\sin^4(\theta/2)\sum_{n=0}^{N+1}r^n$, where we have included a further 
factor $\sin(\theta/2)$ taking into account that $\Delta z$ becomes $c\tau \sin(\theta/2)$
in this configuration. This result can also be obtained in more elegant 
and rigorous way by making the computation in the reference system 
in which the total momentum is zero and the two colliding photons 
are antiparallel. In fact, by indicating with a prime the quantities
in such a system, and being $z$ and $y$ the vertical and horizontal
directions in thelaboratory system
of figure 1, 
we have: $t'=\gamma(t-\beta y/c)$, $y'=\gamma(y-\beta c t)$, 
$z'=z$, $\omega'=\gamma(\omega-\beta c k_y)$,
and $k_y'=\gamma(k_y-\beta \omega/c)\equiv 0$ and $k_z'=k_z$, 
where $\beta=ck_y/\omega$ and 
$\gamma=\sqrt{1-\beta^2}$. It is then easy to see that the phase
$k_z' z'-\omega' t'+2 \epsilon_0{\cal A}^2 (k_z')^3\xi_L \Delta z'$, when 
translated to the laboratory system, gives $k_z z+k_y y-\omega t+\Delta\phi$,
with $\Delta\phi=\sin^4(\theta/2) \Delta\phi_{\theta=\pi}$. 
A similar result can be obtained in the case of orthogonally polarized waves.

Taking $N=1000$ in the expression of $F$ that we have obtained above, 
we can find a limiting value $F_{\rm max}\simeq 50$, 
that in principle can be achieved with present technology.
As a result, the measurement of the phase shifts in the experiments 1) and 2)
that we proposed above with such a gain factor
will be able to resolve $\xi_L$ and $\xi_T$ as small as 
$3.0\times 10^{-30}m^3/J$ and $1.7\times 10^{-30}m^3/J$ respectively, thus allowing to
detect PPS as predicted by QED, or find a signal of non-standard physics.
Note that the combination of the two experiments will also be able to test 
Born Infeld theory and scenarios involving MCPs or axion-like particles,
taking into account the discussion of section 2.

\section{Conclusions}

We have computed the phase shifts affecting a low power laser beam that crosses an
high power laser pulse with general transverse polarisation, and proposed a 
set of experiments to completely determine the parameter space of the 
effective Lagrangian that describes PPS well below the
threshold for the creation of electron-positron pairs.

%%%%%%%%%%%%%%%%%%%%%% FIGURE %%%%%%%%%%%%%%%%%%%
%\begin{figure}[htb]
%{\centering \resizebox{8.5cm}{!}{\includegraphics{log_plot.eps}} \par}
%\caption{Exclusion plot for the search of PPS. 
%The parameters $\xi_L$ and $\xi_T$ are measured in units of $m^3/J$.
%The diagonal line corresponds
%to Born Infeld theory, while the point is the QED prediction.
%The darkest region is excluded by the current PVLAS constraint. 
%The next two inner regions correspond 
%to the parts of the parameter space that can be probed 
%at HERCULES with single crossing ($F=1$) or multiple crossing 
%(choosing $F=30$), respectively.
%Finally, the last inner region represents the range that 
%can be reached at ELI with single crossing.}
%\label{fig2}
%\end{figure}
%%%%%%%%%%%%%%%%%%%%%%%%%%%%%%%%%%%%%%%%%%%%%%%

\FIGURE[p]{
\epsfig{file=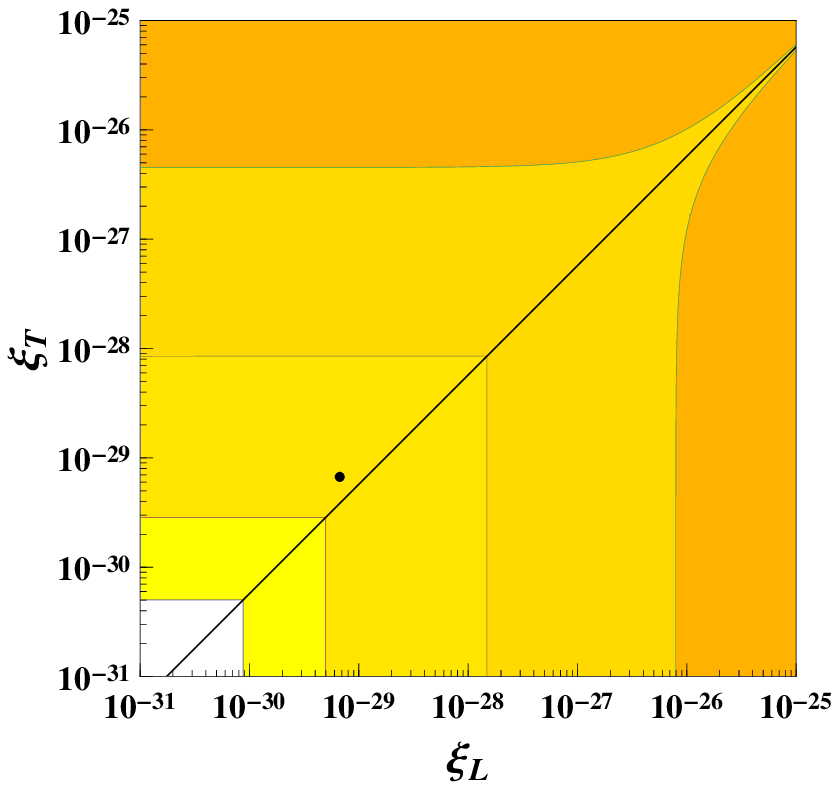}
\caption{Exclusion plot for the search of PPS.  The parameters $\xi_L$
  and $\xi_T$ are measured in units of $m^3/J$.  The diagonal line
  corresponds to Born Infeld theory, while the point is the QED
  prediction.  The darkest region is excluded by the current PVLAS
  constraint.  The next two inner regions correspond to
  the parts of the parameter space that can be probed at HERCULES with
  single crossing ($F=1$) or multiple crossing (choosing $F=30$),
  respectively.  Finally, the last inner region represents the range
  that can be reached at ELI with single crossing.}
} 

Our results are summarised in figure 2, showing the 95\% C.L. exclusion
regions that can be obtained with this set of experiments at
the present facility HERCULES, without or with multi-crossing, as
compared to the current constraint by PVLAS.  The predictions of QED
and Born-Infeld theory are explicitly indicated.  Additional
contributions from minicharged particles, or axion-like scalar 
or pseudoscalar bosons, can sum with
them and produce a different point in the $\xi_L$ and $\xi_T$ plane,
as discussed in section 2.

Even with the single-crossing version ($F=1$), 
figure 2 shows how the PVLAS limits can be substantially
improved at HERCULES, possibly allowing for the detection of PPS of non-standard origin.
On the other hand, by using a multi-cross mechanism, 
HERCULES would already be able to detect PPS of QED origin. 
Note that the result of figure 2 is obtained using the conservative value $F=30$, 
corresponding to just $N=44$ aligned mirrors, 
which is a more realistic assumption than the maximum value that we have found
above. Finally, in figure 2, we also
see how the sensitivity will be improved at ELI in the future, thus
allowing for a more precise determination of the parameters. 

We think that this proposal can eventually contribute to a
new class of precision tests of QED and non-standard models, 
such as Born-Infeld Theory or scenarios involving minicharged particles
or axion-like, scalar or pseudoscalar bosons.

\section{Acknowledgements}

We thank Eduard Masso for explaining us the contraints on axion-like particles, 
and Frank Wise and Guido Zavattini for 
clarifying the sensitivities that can be achieved
in the measurements of phase shifts and ellipticities, respectively. 
We are also grateful to Heinrich Hora and Luis Plaja for helpful remarks
and to Bernardo Adeva for useful discussions.
D.T. thanks Pedro Fern\'andez de C\'ordoba and the 
whole InterTech and IMPA Groups for 
their very nice hospitality at the Universidad Polit\'ecnica de Valencia. 
This work was partially supported by 
Xunta de Galicia, by contract No.
PGIDIT06PXIB239155PR, and by the Government of Spain,
by contracts N. TIN2006-12890, FIS2007-29090-E and FIS2008-0100.

\end{document}